\documentclass[a4paper,11pt]{article}

\usepackage{amsfonts}
\usepackage{amsmath}
\usepackage{amssymb}
\usepackage{epsfig}

\def\beq{\begin{equation}}
\def\eeq{\end{equation}}
\def\bea{\begin{eqnarray}}
\def\eea{\end{eqnarray}}

\def\u1{\widehat{U(1)}}
\def\su2{\widehat{SU(2)}_1}

\def\uq2{U_q(sl(2))}
\def\Uq2{\widehat{U_q(sl(2)}}

\def\a{\alpha}
\def\b{\beta}

\def\l{\lambda}

\def\de{\partial}

\def\ket{\rangle}

\begin{document}

\begin{titlepage}

\begin{center}
\hfill  \quad  \\
\vskip 1 cm
{\Large \bf  Critical Theory  of Two-Dimensional Mott Transition:
Integrability and Hilbert Space Mapping}

\vspace{1cm}

\vspace{.4cm}
Federico~L.~ BOTTESI$^1$ ,\ \ Guillermo~R.~ZEMBA$^{1,2}$\\

\bigskip
{\em $^1$Facultad de Ingenier\'ia  Pontificia Universidad Cat\'olica Argentina,}\\
{\em  Av Alicia Moreau de Justo 1500, 1428, Buenos Aires, Argentina\\}
\medskip
{\em $^2$Physics Department,}\\
{\em  Comisi\'on Nacional de Energ\'{\i}a At\'omica,} \\
{\em Av.Libertador 8250, (1429) Buenos Aires, Argentina}

\end{center}
\vspace{.5cm}
\begin{abstract}
\noindent
We reconsider the Mott transition in the context of a two-dimensional fermion model with density-density coupling. 
We exhibit a Hilbert space mapping between the original
model and the Double Lattice Chern-Simons theory at the critical point by
use of the representation theory of the $q$-oscillator and Weyl algebras . 
The transition is further characterized by the ground state modification. 
The explicit mapping provides a new tool to further
probe and test the detailed physical properties of the fermionic lattice model considered here
and to enhance our understanding of the Mott transition(s).
\end{abstract}

\vskip 1.cm


\end{titlepage}
\pagenumbering{arabic}

\noindent{\bf Introduction}
\vskip 3mm
The physical properties of strongly correlated electron systems are difficult to predict or even to describe,  
mainly because of the lack of suitable reliable tools to study them. 
Among these systems, the Mott Insulators ({\it i.e.}, electronic systems which undergo a metal-insulator transition driven by correlations) have a prominent place.  
Most of the studies of the Mott transition are based on the microscopic dynamics of the electron
system. The models are defined by a electron Hamiltonian that is then solved either by 
some approximation or by numerical methods. Both methods have their limitations, which have 
been discussed, {\it e.g.}, in \cite{Botte-Z-Mott1}. 
As of today, there are two basic tools to study the Mott transition that complement themselves. One is the Dynamical Mean Field Theory (DMFT) method, 
valid in the limit of infinite space dimensions \cite{DMFT} . The other tool is
the use of integrability properties such as the Bethe Ansatz or bosonization techniques in some specific
models, mostly in one spatial dimension.
Different, non-perturbative approaches 
to models that display some form of Mott transition are therefore desirable. One such approach is that 
of the EFTs \cite{Polch}, which have been shown to be a powerful tool for dealing with 
strongly correlated systems (in particle, condensed matter physics and statistical mechanics,
among other areas of knowledge). The EFT in condensed matter physics has its roots in
Landau's ideas of effective degrees of freedom and their characterization throughout
symmetry. It has been further developed after the introduction of the Renormalization
Group ideas following Wilson's approach. The main idea is to first identify the characteristic
effective degrees of freedom of a given system at a given energy scale (usually arising 
from the phenomenology), identifying their symmetries  and then writing down the most general 
Hamiltonian (or Lagrangian) compatible with those symmetries. 

For the case of the Mott transition, we have applied the EFT method to a fermion model on the
lattice with density-density coupling in a previous paper \cite{Botte-Z-Mott1}. In that article, we have provided an extension of the method of integrability to a $(2+1)$-dimensional spinless fermion model with nearest neighbors Coulomb interactions, having written down an Effective Field Theory (EFT) to further study the properties of the model at the Mott transition critical point. 
The goal of the present article is to reformulate this approach in a different, perhaps more straightforward
fashion which could be useful for future developments and generalizations,
and to shown that the EFT previously obtained is actually the corresponding (equivalent ??) field theory at the level of the Hilbert space at the critical point. 
Moreover, under this approach, we will shown that, the Mott transition is characterized as a change in the ground state.

\vskip 3mm
\noindent{\bf Fermionic model and its integrability}
\vskip 3mm
We start by considering the following Hamiltonian model :

\beq
H_{2D}\ =\ -\frac{t}{2}\ \sum_{x,\mu} [\ \psi^\dagger(x+ a e_\mu) e^{i A_\mu} \psi(x)\ +\ {\rm h.c.}\ ]\  +\ U\ \sum_{x,\mu} \rho(x)\rho(x+a e_\mu)\ , \label{Model-Ferm-2d}
\eeq
where $\psi(x)$ is the fermion field, $x$ labels the lattice sites and $e_\mu$ are the unit lattice vectors pointing to the nearest neighbors of a given site, $a$ is the lattice spacing ,
$t$ is the hopping parameter, $U$ is the (constant) Coulomb potential, $\rho(x)$ is the  charge density (normal-ordered with respect to the half-filling ground state),  $\rho(x)= [:\psi^{\dagger}(x)\psi(x): -1/2]$ and $A_\mu $
is an Abelian statistical gauge field defined on the links of the lattice.

This model can be mapped into the two-dimensional anisotropic Heisenberg ($XXZ$ spin) model by means of a two-dimensional Jordan-Wigner transformation \cite{Fradkin} :

\bea
&& S^+(x)=\psi^\dagger (x) U_{2D} \\
&& S^-(x)=  U^\dagger_{2D}  \psi(x) \\
&& S_z(x)=\psi^\dagger(x) \psi(x)-1/2\\
&& U_{2D}(x)=e ^{ i \sum_{x,y}\Theta(x,y) \psi^{\dagger}_j\psi_(y)} \\
&& A_{\mu}(x)=\sum_k [\Theta(k,x)-\Theta(k,x+\epsilon_{\mu}] \psi^\dagger_k \psi_k 
\eea
where $S^+(x)$ and $S^-(x)$ are the rising and lowering spin operators for spin one-half particles, and $\Theta(x,y) $ is the lattice angle between two points in a two-dimensional square lattice.

The partition function of the two-dimensional Heisenberg model, in the Hamiltonian framework, can be written as:

\beq
Z=Tr_{H_\gamma V_x V_y}[T(x_1,y_1)......T(x_n,Y_n)]\ , \eeq

where $H_\gamma$ is the Quantum (Hilbert) space , $V_x=\otimes_{i} V_{xi}$ is the row-space , $V_y=\otimes_{i} V_{yi}$ is the column-space, and $T_{(x_iy_j)}(u)$ 
is the layer-to-layer transfer matrix given by:or 

\bea
&&  T_{(x_iy_j)}(u)=\exp(uH_{xxz}(x_i,y_j)) \\
&& H_{XXZ}\ =\ \sum_{\langle i j\rangle}\left [ (\ S^x_i S^x_j\ +\ S^y_i S^y_j\ ) -\Delta S^z_i S^z_j \right]
\eea
where $\Delta=-t/U$. The integrability of the model requires the commutativity of the layer-to-layer transfer matrices, which is guaranteed by the existence 
of solutions of the Zamolodchikov Tetrahedron equation (ZTE)(\cite{Zamolodchikov-ZTE})
  
 \beq
{\bf R}_{V_1,V_2,V_3} {\bf R}_{V_1,V_4,V_5} {\bf R}_{V_2,V_4,V_6} {\bf R}_{V_3,V_5,V_6}=  
{\bf R}_{V_3,V_5,V_6} {\bf R}_{V_2,V_4,V_6} {\bf R}_{V_1,V_4,V_5}  {\bf R}_{V_1,V_2,V_3} \eeq
where, we have made some abuse of notation since now  $V_i$ could be the quantum-Hilbert space  or the row / column sates.   

As it is known, the $R$-matrix provides an intertwining for the layer-to-layer transfer matrix  (or, equivalently, 
for the $L$ operators ), {\it i.e.}, it satisfies:

\beq
  {\bf L}_{ab,1}{\bf L}_{ac,2}{\bf L}_{bc,3}{\bf R}_{123}= {\bf R}_{123}{\bf L}_{bc,3}{\bf L}_{ac,2}{\bf L}_{ab,1} \label{LLLR-RLLL}\ ,\eeq
where the $L$ operators act on the tensor product vector space $ V_{x_i}\otimes V_{x_j}\otimes H_\gamma$. (Here the Latin index a,b,c stands for classical spin-1/2 
representation spaces and numeric index stands for the quantum Hilbert spaces) 
The existence of solutions of the ZTE follows from the solutions of the Quantum Korepanov Equation (QKE) \cite{Korepanov}
 
 \beq
X_{a,b}[A_1]X_{a,c}[A_2]X_ {b,c}[A_3]=X_ {b,c}[A'_3]X_{a,c}[A'_2]X_{a,b}[A'_1]\eeq
which codifies the zero-curvature condition of a 'quantum scattering problem'. Here $A_1$ $(A'_1)$ represent a 
algebra of observables and $X_{\a,\b}$ acts on the direct sum of vector spaces $V_\a\oplus V_\b$.
In an outstanding series of articles, Sergeev et al.\cite{Sergeev-integrability-q-oscilator} \cite{Bazhanov1} have shown that (under minimal conditions) the only solution of the QKE for 'vertex type problems' ( {\it i.e.}, when the lattice problem is formulated in terms of vertex potentials) is given by:
 
 \[X ( O_{1q})  =  \left [ \begin{array}{ccc}
{k_1} & {a^*_1} & {0}           \\
{-a_1} & {k_1} &{ 0}  \\
 {0} & {0} & {1}\  \\
  \end{array}\right]\]
where $O_q$ means that the operators in the Korepanov matrix carry representations of the $q$-oscillator 
algebra, {\it i.e.} they satisfy:
  
\begin{eqnarray}
&& qa^\dagger a-q^{-1}a a^\dagger =q-q^{-1 } \\
&&ka^{\dagger}=q a^{\dagger}k \\
&& ka=q^{-1}ak\\
&& k^2=q(1-a^\dagger a)=q^{-1}(1-aa^·\dagger) \end{eqnarray}
where $q$ is the deformation parameter. The corresponding $L$ operators are:

\bea
L_{\a_i,\b_j}(O_{ij},i,j)=\begin{pmatrix}
  1 & 0   &  0       & 0       \\ 
  0 & \l q^{h_v}  & \nu  a_v^\dagger & 0 \\
   0 & -\nu a_v  &  \mu q^{h_v}     & 0  \\
   0 &  0    & 0         & \nu^2\ 
\end{pmatrix} \eea
where we have used $k=q^{h}=q^{s_z/2}$ and have introduced the afinization parameters $(\mu \nu)$. 
Is straightforward to show that the products of two $L$ operators
of the $q$-oscillator model give rise to an $L$ operator of the Heisenberg $XXZ$ model (for details 
see  \cite{Sergeev-Qos} \cite{Botte-Z-Mott1} ). Therefore, for a square 
lattice with an even number of sites ( on the rows and on the columns) the partition function of the 
$XXZ$ model can be written as:

\beq
Z=Tr_{H_\gamma V_x V_y}[ L(x_1,y_1, O_{1,1})........................L(x_n,y_n,O_{nn}]\ . \eeq
This means that  the  original model is mapped onto the $q$-oscillator model. It then becomes possible 
to study the states of the lattice fermion model (\ref{Model-Ferm-2d})
by analyzing the representations of the $q$-oscillator algebra.

For $q=e^{\zeta h}<1 $ the $q$-oscillator algebra has Fock space representations defined by:
\bea
&& q^{\bf N}|n\ket= q^n|n\ket \\
&&  a^+|n\ket=\sqrt{1-q^{(2n+2)}}|n+1\ket  \quad  a^- |n\ket=\sqrt{1-q^{2n}}|n-1\ket \quad (b^-)^\dagger=b^+ \nonumber \\
&&  a^+|n\ket=\sqrt{q^{(2n)}1}|n-1\ket  \quad  a^- |n\ket=- \sqrt{q^{2n+2}-1}|n+1\ket \quad (b^-)^\dagger=-b^+ \nonumber \\ \eea
for $n\ge 0$ and $n<-1$, respectively.
Furthermore, the states of the system are:
\bea
|\psi \rangle=\otimes_{ij} |n_{ij}\rangle \eea
To achieve a deeper understanding  of the solution that we have just discussed, we may use a crucial property of the ZTE. Namely, the ZTE can be projected ( or reduced) onto the Yang-Baxter equation after tracing 
out over one ( temporal or spatial )  direction. Tracing out over the $y$-column we obtain a one-dimensional Heisenberg $XXZ$ chain, which is known to belong to the universality 
class of the Luttinger liquids, impliying that the degrees of freedom of this chain are charge density waves. 
This observation fits within the picture of the solution  as a 'quantum fluctuation'.
Moreover, since the ZTE can be projected on any row or column , the consistency of the theory demands that the solution must be a two-dimensional charge fluctuation on the lattice. 
Hence, the $q$ parameter becomes a {\it two-dimensional analog of the Luttinger parameter}.

In order to identify the critical point with the values of the parameters in the fermion model (\ref{Model-Ferm-2d}) let us remind that the reduced one-dimensional model
( which has a long history ) have been solved in \cite{Shankar} , and it is known to undergo a metal-insulator (Mott transition), and a charge density wave ordering
(CDW) with a breakdown of the parity symmetry above the Mott gap. This Mott transition appears when $t=U$, {\it i.e.}, when the dimensionless parameter 
$\Delta=-t/U=-(q+q^{-1})/2=1$  
.


We will now study the representations of the q-Oscillattor algebra at the Mott transition point $q=-1$. When the deformation parameter satisties  $q^2=1$,
the algebra reduces  to a two independent Weyl algebras:
\beq
W_q : kb^{+}=q b^{+} k  \quad W_{q^{-1}} : kb^{-}=1/q b^{+} k  \qquad \{b^+,b^-\}=0 \eeq
which have cyclic representations for $q^{2N}=1$, $q^N=-1$ given by:
\beq
k|m \rangle =q^{m} |m\rangle \qquad b|m\rangle=|m+1\rangle .\eeq

Now we claim that the 'corresponding' field theory at the level of the Hilbert space is a Double-Lattice-Chern Simons theory with abelian gauge group.
To show this, first we shall  impose periodic boundary conditions in the original fermion model, and compactify the manifold onto a torus such that 
the original square lattice matches with the lattice made by the homology cycles of this torus, and consider the Abelian C-S action  \cite{Bos-Nair}

\beq
S=\frac{4\pi}{k} \int d^3x \epsilon^{\mu,\nu,\lambda} A_\mu \de_\nu A_\lambda \ . \eeq
this a topological gauge field theory with natural observables provided by Wilson Loops:
\beq
W_\gamma=Pexp{(i\oint _{\gamma} A dl)}\ .  \eeq
In holomorphic coordinates, the gauge field may be decomposed as:
\beq
A_{\bar{z}}=\de_{\bar{z}} \chi + \frac{i\pi}{Img(\tau)}\bar{\omega}(z) a\ , \eeq
where $\tau$ is the modular parameter of the torus, and $\omega_i$, $\bar{\omega_i}$ is a basis of 
holomorphic $1$-forms on the torus , $a(t)$ is a complex parametric function.
The wave functional may be written as:
\beq
\Psi[A]=\psi(\chi)\psi(a)\label{basis-holo}  . \eeq

The local Gauge transformations on the CS theory are defined by: $U(x)=g(\alpha(x))=exp(i\a (x))$. 
However, on the torus the gauge theory may also have  
global gauge transformations associated with the windings (of the Wilson loops) over the non-contractible loops around the torus. 
Let us denote by $U_{n,m}$ the gauge transformations with
$n$ and $m$ integer winding numbers around the (orthogonal) homology  cycles. These global gauge transformations have anomalous commutations
relations  that can be avoided by requiring the condition:
(for details see \cite{grensing}):
\beq
U_{(n,0)}\psi(a)=e^{2i\pi n.\mu}\psi(a) \qquad U_{(0,m)}\psi(a)=e^{2i\pi n.\nu}\psi(a)\ ,  
\eeq
where $\nu_i$ $\mu_i$ are parameters that belong to the interval $[0,1]$. These conditions are solved by the 
Jacobi Theta functions with solutions labeled by an integer $m=1,2...k$.
The large Gauge transformations still act as symmetries of the Chern-Simons theory and a  basis of such Gauge transformations  
may be written in terms of the Wilson loops as: 
\beq
U_{(1,0)}=e^{i\int_{C_x} A}\equiv S \qquad  U_{(0,1)}=e^{i\int_{C_y}A} \equiv T\ \label{S-Q-operators}, \eeq
which satisfy a Weyl-algebra:
\beq
ST=qTS \eeq

On the basis 
\beq 
\Psi_m(A_z)=\langle m |\Psi\rangle \label{basis-CS2}  \eeq
the operators(\ref{S-Q-operators})  act as:
\bea 
&& S_i|m\rangle =q^{m_i+\mu_i} |m\rangle  \\
&&  T_i|m\rangle=q^{\nu_i}|......m_i-1,......\rangle
\label{CS-states}\eea
where  $q=e^{i\pi/k}$.  

Then, taken $\mu_i=\nu_i=0$ ({\it i.e.}, using bosonic boundary condition for the CS-field) and identifying $S \rightarrow k$ and $T \rightarrow b^+$ 
we see that the states of the $q$-oscillator (and therefore the sates 
of the Fermion model) at  $q=-1$ correspond to the states of the CS Theory (\ref{basis-holo})
in the basis (\ref{basis-CS2}).
The coupling constant  of the CS theory may now be inferred in two different ways: 
Firstly, we note that at the Mott transition:
\beq
\Delta=1 \Rightarrow q=-1 \Rightarrow k=1\ .\eeq
Secondly, we note that the projection property of the ZTE implies that each row (or column) is a $XXZ$ spin chain (whose critical properties are described by a Weiss-Zumino-Witten model with coupling constant $k=1$), which is known to match the coupling constant of the corresponding $(2+1)$ CS theory ($k=1$). For further details, please see our analysis in  \cite{Botte-Z-Mott1}).

Taking into account that the degrees of freedom of the Fermion Model (\ref{Model-Ferm-2d}) must be restricted to a square lattice, using the fact that the $q$-oscillator algebra splits into two 
Weyl algebras and using the parity of the original model, we deduce that the corresponding (equivalent) Field Theory at 
the Mott critical point of the fermion model (\ref{Model-Ferm-2d}) is a 
Double Lattice Chern-Simons theory:

\bea
S_{DCS}=\frac{k}{4\pi} \int d^3x\ a^R_\mu K_{\mu,\nu} a^R_{\nu} -\frac{k}{4\pi} \int d^3x\  a^L_\mu K_{\mu,\nu}a^L_{\nu}\ , \label{Double-CS}
\eea 
with coupling constant $k=1$, where $a^R$ and $a^L$ are two Abelian gauge fields of opposite chiriality (left and right), and  where $K_{\mu,\nu}=S_{mu}\epsilon_{\mu,\a,\nu}d_\a$, $S_\mu f(x)=f(x+a\epsilon_\mu)$, $d_\mu f(x)=(f(x+a\epsilon_\mu)-f(x))/a$, (where  $a$  is the lattice-parameter) \cite{Carlo-Topics}.
This theory has quantum group symmetry $U_q(\widehat{sl(2)})\otimes U_q(\widehat{sl(2)})$ with deformation parameter $q=-1$ \cite{Witten-2} \cite{grensing}.

\vskip 3mm
\noindent{\bf Conclusions}
\vskip 3mm

In this article we have reconsidered the integrability of the two-dimensional density-density coupled fermion moldel  (\ref{Model-Ferm-2d}),  which follows from the 
solution of the Zamolodchikov's Tetrahedron equation associated with the q-oscillator algebra, firstly found in (\cite{Bazhanov1}). 
Using the representation theory of this algebra, we have constructed a explicit mapping between the states of the original fermion model at the Mott critical point ($\Delta=-1$) and the states 
of the lattice Double Chern Simons theory  at coupling constant $k=1$. 
That is, we have provided an explicit link between the Hilbert spaces of a microscopic
theory with those of its long distance EFT, something that can not be expected in general
systems, although some researchers in the condensed matter community frequently ask for. 
The changes in the representation theory of the q-oscillator algebra signal a change in the ground states of the Fermion Model and provide us a tool to further investigate this ground
state transition. 
The significance of the explicit mapping is that it provides a new tool to further
probe and test the detailed physical properties of the fermionic lattice model considered here.
In our previous work, some of the correct long-distance physical properties of this model
(predicted by the EFT) were somehow hidden in the intricacy of the mappings among the different 
models and theories used to establish the equivalence of them. The ability to provide more explicit 
answers to interesting questions arising in the context of this fermionic model gives us
hope to use it as a tool to further develop our understanding of the nature of the Mott transition(s). 
Another goal we had in mind writing this paper was to provide a further link between the
three different areas of research common to the type of systems considered here, namely: condensed 
matter, theoretical and mathematical physics.
 \vskip 3mm

\noindent{\bf Acknowledgments}

 \vskip 3mm
 \noindent
G. R. Zemba is a member of CONICET (Argentina).

\def\RMP{{\it Rev. Mod. Phys.\ }}

 \def\PRL{{\it Phys. Rev. Lett.\ }}
 \def\CMP{{\it Commun.Math.Phys.\ }}
 \def\PL{{\it Phys. Lett.}}

 \def\PR{{\it Phys. Rev.  \ }}
 \def\NP{{\it Nuclear. Phys.}}
 \def\PRB{{\it Phys. Rev. B  \ }}
 \def\IJMP{{\it Int. J. Mod. Phys.}}

\end{document}